\documentclass[twocolumn,prl,amsmath,amssymb,superscriptaddress]{revtex4-2}

\usepackage{lineno,hyperref}
\usepackage{amssymb}
\usepackage{color}
\usepackage{epstopdf}
\usepackage{graphicx}
\usepackage{dcolumn}
\usepackage{bm}
\usepackage{multirow}
\usepackage{amsmath}
\usepackage{longtable}
\usepackage{epstopdf}
\usepackage{booktabs}

\begin{document}

\title{First Exploration of Monopole-Driven Shell Evolution above the $\mathbf{N=126}$ shell closure: new Millisecond Isomers in $^{213}$Tl and $^{215}$Tl}%

\author{T.T.~Yeung}
\affiliation{Department of Physics, The University of Tokyo, 7-3-1 Hongo, Bunkyo, Tokyo 113-0033, Japan}
\affiliation{RIKEN Nishina Center, 2-1 Hirosawa, Wako, Saitama 351-0198, Japan}

\author{A.I.~Morales}
\email{Ana.Morales@ific.uv.es}
\affiliation{Instituto de Física Corpuscular, CSIC-Universitat de Val\`encia, E-46071 Val\`encia, Spain}

\author{J.~Wu}
\email{jwu2@bnl.gov}
\affiliation{National Nuclear Data Center, Brookhaven National Laboratory, Upton, New York 11973, USA}
\affiliation{Facility for Rare Isotope Beams, Michigan State University, East Lansing, Michigan 48824, USA}

\author{M.~Liu}
\affiliation{Sino-French Institute of Nuclear Engineering and Technology, Sun Yat-Sen University, Zhuhai, Guangdong 519082, China}

\author{C.~Yuan}
\affiliation{Sino-French Institute of Nuclear Engineering and Technology, Sun Yat-Sen University, Zhuhai, Guangdong 519082, China}

\author{S.~Nishimura}
\affiliation{RIKEN Nishina Center, 2-1 Hirosawa, Wako, Saitama 351-0198, Japan}

\author{V. H. Phong}
\affiliation{RIKEN Nishina Center, 2-1 Hirosawa, Wako, Saitama 351-0198, Japan}
\affiliation{Faculty of Physics, VNU Hanoi University of Science, 334 Nguyen Trai, Thanh Xuan, Hanoi, Vietnam}

\author{N.~Fukuda}
\affiliation{RIKEN Nishina Center, 2-1 Hirosawa, Wako, Saitama 351-0198, Japan}

\author{J.L.~Tain}
\affiliation{Instituto de Física Corpuscular, CSIC-Universitat de Val\`encia, E-46071 Val\`encia, Spain}

\author{T.~Davinson}
\affiliation{School of Physics and Astronomy, University of Edinburgh, EH9 3FD Edinburgh, UK}

\author{K.P.~Rykaczewski}
\affiliation{Physics Division, Oak Ridge National Laboratory, Physics Division, TN 37831-6371, USA}

\author{R.~Yokoyama}
\affiliation{Department of Physics and Astronomy, University of Tennessee, Knoxville, Tennessee 37996, USA}
\affiliation{Center for Nuclear Study, the University of Tokyo, 2-1 Hirosawa, Wako Saitama 351-0198, Japan}

\author{T.~Isobe}
\affiliation{RIKEN Nishina Center, 2-1 Hirosawa, Wako, Saitama 351-0198, Japan}

\author{M.~Niikura}
\affiliation{RIKEN Nishina Center, 2-1 Hirosawa, Wako, Saitama 351-0198, Japan}

\author{Zs.~Podoly\'ak}
\affiliation{Department of Physics, University of Surrey, Guildford, Surrey GU2 7XH, United Kingdom}

\author{G.~Alcal\'a}
\affiliation{Instituto de Física Corpuscular, CSIC-Universitat de Val\`encia, E-46071 Val\`encia, Spain}

\author{A.~Algora}
\affiliation{Instituto de Física Corpuscular, CSIC-Universitat de Val\`encia, E-46071 Val\`encia, Spain}
\affiliation{HUN-REN Institute for Nuclear Research, Bem t\'er 18/c, Debrecen H4032, Hungary}

\author{J.~Agramunt}
\affiliation{Instituto de Física Corpuscular, CSIC-Universitat de Val\`encia, E-46071 Val\`encia, Spain}

\author{C.~Appleton}
\affiliation{School of Physics and Astronomy, University of Edinburgh, EH9 3FD Edinburgh, UK}

\author{H.~Baba}
\affiliation{RIKEN Nishina Center, 2-1 Hirosawa, Wako, Saitama 351-0198, Japan}

\author{R. Caballero-Folch}
\affiliation{TRIUMF, Vancouver BC, V6T 2A3, Canada}

\author{F.~Calvino}
\affiliation{Universitat Politecnica de Catalunya, E-08028 Barcelona, Spain}

\author{M.P.~Carpenter}
\affiliation{Physics Division, Argonne National Laboratory, Argonne, IL 60439, USA}

\author{I.~Dillmann}
\affiliation{TRIUMF, Vancouver BC, V6T 2A3, Canada}
\affiliation{Department of Physics and Astronomy, University of Victoria, Victoria, BC V8P 5C2, Canada}

\author{A.~Estrade}
\affiliation{Department of Physics, Central Michigan University, Mount Pleasant, MI, 48859, USA}

\author{T.~Gao}
\affiliation{Department of Physics, The University of Hong Kong, Pokfulam, Hong Kong, China}

\author{C.J.~Griffin}
\affiliation{TRIUMF, Vancouver BC, V6T 2A3, Canada}

\author{R.~Grzywacz}
\affiliation{Department of Physics and Astronomy, University of Tennessee, Knoxville, Tennessee 37996, USA}
\affiliation{Physics Division, Oak Ridge National Laboratory, Physics Division, TN 37831-6371, USA}

\author{O.~Hall}
\affiliation{School of Physics and Astronomy, University of Edinburgh, EH9 3FD Edinburgh, UK}

\author{Y.~Hirayama}
\affiliation{Wako Nuclear Science Center (WNSC), Institute of Particle and Nuclear Studies (IPNS), High Energy Accelerator Research Organization (KEK), Wako, Saitama 351-0198, Japan}

\author{B.M.~Hue}
\affiliation{Institute of Physics, Vietnam Academy of Science and Technology, Hanoi 10000, Vietnam}

\author{E.~Ideguchi}
\affiliation{Research Center for Nuclear Physics (RCNP), Osaka University, 10-1 Mihogaoka, Ibaraki, Osaka 567-0047, Japan}

\author{G.G.~Kiss}
\affiliation{HUN-REN Institute for Nuclear Research, Bem t\'er 18/c, Debrecen H4032, Hungary}
\affiliation{RIKEN Nishina Center, 2-1 Hirosawa, Wako, Saitama 351-0198, Japan}

\author{K.~Kokubun}
\affiliation{Department of Physics, The University of Tokyo, 7-3-1 Hongo, Bunkyo, Tokyo 113-0033, Japan}

\author{F.G.~Kondev}
\affiliation{Physics Division, Argonne National Laboratory, Argonne, IL 60439, USA}

\author{R.~Mizuno}
\affiliation{Department of Physics, The University of Tokyo, 7-3-1 Hongo, Bunkyo, Tokyo 113-0033, Japan}

\author{M.~Mukai}
\affiliation{Department of Applied Energy, Nagoya University, Nagoya 464-8603, Japan}

\author{N.~Nepal}
\affiliation{Department of Physics, Central Michigan University, Mount Pleasant, MI, 48859, USA}

\author{M.N. Nurhafiza}
\affiliation{Department of Physics, Osaka University, Machikaneyama-machi 1-1, Osaka 560-0043 Toyonaka, Japan}

\author{S.~Ohta}
\affiliation{Center for Nuclear Study, the University of Tokyo, 2-1 Hirosawa, Wako Saitama 351-0198, Japan}

\author{S.E.A.~Orrigo}
\affiliation{Instituto de Física Corpuscular, CSIC-Universitat de Val\`encia, E-46071 Val\`encia, Spain}

\author{M.~Pall\`as}
\affiliation{Universitat Politecnica de Catalunya, E-08028 Barcelona, Spain}

\author{J.~Park}
\affiliation{Center for Exotic Nuclear Studies, Institute for Basic Science, Republic of Korea}

\author{D.~Rodr\'iguez-Garc\'ia}
\affiliation{Instituto de Física Corpuscular, CSIC-Universitat de Val\`encia, E-46071 Val\`encia, Spain}

\author{H.~Sakurai}
\affiliation{RIKEN Nishina Center, 2-1 Hirosawa, Wako, Saitama 351-0198, Japan}
\affiliation{Department of Physics, The University of Tokyo, 7-3-1 Hongo, Bunkyo, Tokyo 113-0033, Japan}

\author{L.~Sexton}
\affiliation{School of Physics and Astronomy, University of Edinburgh, EH9 3FD Edinburgh, UK}
\affiliation{TRIUMF, Vancouver BC, V6T 2A3, Canada}

\author{Y.~Shimizu}
\affiliation{RIKEN Nishina Center, 2-1 Hirosawa, Wako, Saitama 351-0198, Japan}

\author{H.~Suzuki}
\affiliation{RIKEN Nishina Center, 2-1 Hirosawa, Wako, Saitama 351-0198, Japan}

\author{A.~Sveiczer}
\affiliation{HUN-REN Institute for Nuclear Research, Bem t\'er 18/c, Debrecen H4032, Hungary}

\author{H.~Takeda}
\affiliation{RIKEN Nishina Center, 2-1 Hirosawa, Wako, Saitama 351-0198, Japan}

\author{A.~Tarifeno-Saldivia}
\affiliation{Universitat Politecnica de Catalunya, E-08028 Barcelona, Spain}
\affiliation{Instituto de Física Corpuscular, CSIC-Universitat de Val\`encia, E-46071 Val\`encia, Spain}

\author{A.~Tolosa-Delgado}
\affiliation{Department of Physics, University of Jyvaskyl\"a, 40014 Jyv\"askyl\"a, Finland}

\author{J.A.~Victoria}
\affiliation{Instituto de Física Corpuscular, CSIC-Universitat de Val\`encia, E-46071 Val\`encia, Spain}

\author{Y.X.~Watanabe}
\affiliation{Wako Nuclear Science Center (WNSC), Institute of Particle and Nuclear Studies (IPNS), High Energy Accelerator Research Organization (KEK), Wako, Saitama 351-0198, Japan}

\author{J.M. Yap}
\affiliation{Department of Physics, The University of Hong Kong, Pokfulam, Hong Kong, China}

\date{\today}%

\begin{abstract}
	
Isomer spectroscopy of heavy neutron-rich nuclei beyond the $N=126$ closed shell has been performed for the first time at the Radioactive Isotope Beam Factory of the RIKEN Nishina Center. New millisecond isomers have been identified at low excitation energies, 985.3(19) keV in $^{213}$Tl and 874(5) keV in $^{215}$Tl. The measured half-lives of 1.34(5) ms in $^{213}$Tl and $3.0(3)$ ms in $^{215}$Tl suggest spins and parities $11/2^-$ with the single proton-hole configuration $\pi h_{11/2}$ as leading component. They are populated via $E1$ transitions by the decay of higher-lying isomeric states with proposed spin and parity $17/2^+$, interpreted as arising from a single $\pi s_{1/2}$ proton hole coupled to the $8^+$ seniority isomer in the $^{A+1}$Pb cores. The lowering of the $11/2^-$ states is ascribed to an increase of the $\pi h_{11/2}$ proton effective single-particle energy as the second $\nu g_{9/2}$ orbital is filled by neutrons, owing to a significant reduction of the proton-neutron monopole interaction between the $\pi h_{11/2}$ and $\nu g_{9/2}$ orbitals. The new $ms$\textendash isomers provide the first experimental observation of shell evolution in the almost unexplored $N>126$ nuclear region below doubly-magic $^{208}$Pb.

\end{abstract}

\maketitle

The study of nuclei far from the valley of stability, so-called exotic or rare nuclei, is crucial to understand the evolution of the shell-model orbitals at extreme numbers of protons ($Z$) and neutrons ($N$) \cite{Thi75,Oza00,Nav00,Bas07,Ste13,Wie13}. The resulting structural properties in rare nuclei \cite{Doo13,Tsu14,Kre16,Now16,Tan19,Ots20} are determinant to advance our knowledge of the reaction and decay rates in key astrophysical events \cite{Wan21,Ali22,Suz22}, in particular the nucleosynthesis processes producing elements heavier than iron \cite{Mum16,Cow21}. In this regard, systematic studies of nuclear states described by single-particle excitations provide invaluable insight on the evolution of shell structure near closed shells. This is a key aspect to better predict structural phenomena in yet unexplored nucleosynthetic nuclei \cite{Now16,Tan20}.

Hitherto, the heaviest spherical closed shells known are $Z=82$ and $N=126$, constituting doubly-magic $^{208}_{82}$Pb$_{126}$.
To what extent the heaviest shell gaps hold away from $^{208}$Pb and what their impact is on the timescales and final abundances of the rapid neutron-capture (\textit{r}) process \cite{Hol23} still remains a conundrum. The weak interaction rates in this region are particularly crucial since they regulate the nucleosynthesis of the actinides via the abundance stockpiles at the third $r$-process waiting point \cite{Nis16,Mum16,Ben12,Kur14,Mor14,Cab16,Car20,Hol23}. They strongly depend on the ordering and energies of single proton and neutron orbitals near the Fermi surface, the most relevant of which have large total angular momentum $j$ (see Fig. 2 of Ref. \cite{Gra07}, pp. 1529). In many theoretical frameworks \cite{Bor03C,Fan13,Mar16}, the $\beta$ decay  of $N\sim126$ nuclei is dominated by high-energy parity-changing first-forbidden (ff) transitions that prevail over the allowed Gamow-Teller (GT) decays $(\nu 0h_{9/2}\rightarrow \pi 0h_{11/2})$ and $(\nu 0i_{11/2}\rightarrow \pi 0i_{13/2})$, resulting in substantially enhanced rates. Since the contributions of ff and GT transitions are affected by changing single-particle energies (SPEs), shell evolution around $^{208}$Pb, specially of the high-$j$ orbitals driving GT strength, becomes of general interest in both nuclear structure and nuclear astrophysics.

In the present Letter, we report on the development of spin isomerism in the single proton-hole $\pi 0h_{11/2}$ states of $^{213}$Tl$_{132}$ and $^{215}$Tl$_{134}$, with six and eight neutron particles above $N=126$. These are the first millisecond isomers reported in the $N>126$ region below $^{208}$Pb, and might prelude the development of astromers that impact on the final elemental abundances of heavy nuclei in the \textit{r} process \cite{Wen21}. 
The new $J^{\pi}=(11/2^-)$ levels have energies about 300 keV lower than the ones predicted by shell-model calculations \cite{Yua22}. The energy decrease is ascribed to an upward trend of the $\pi 0h_{11/2}$ proton effective SPE (ESPE) as the $\nu 1g_{9/2}$ shell is filled by neutrons. This effect is driven by the monopole interaction, which is the lowest order component of the residual nucleon\textendash nucleon interaction \cite{Gra07}.
The results thus provide the first experimental observation of shell evolution in a high-$j$ orbital beyond $^{208}$Pb, 
a key structural feature to help improving the global nuclear models used in \textit{r}-process simulations \cite{Mol03,Mar16,Ney20}. These, at present, show strong discrepancies towards the $N=126$ waiting point and beyond, and need stringent constraints to provide higher-quality calculated $\beta$-decay inputs getting far away from stability.

 \begin{figure*}
	\centering
	\includegraphics[width=\textwidth]{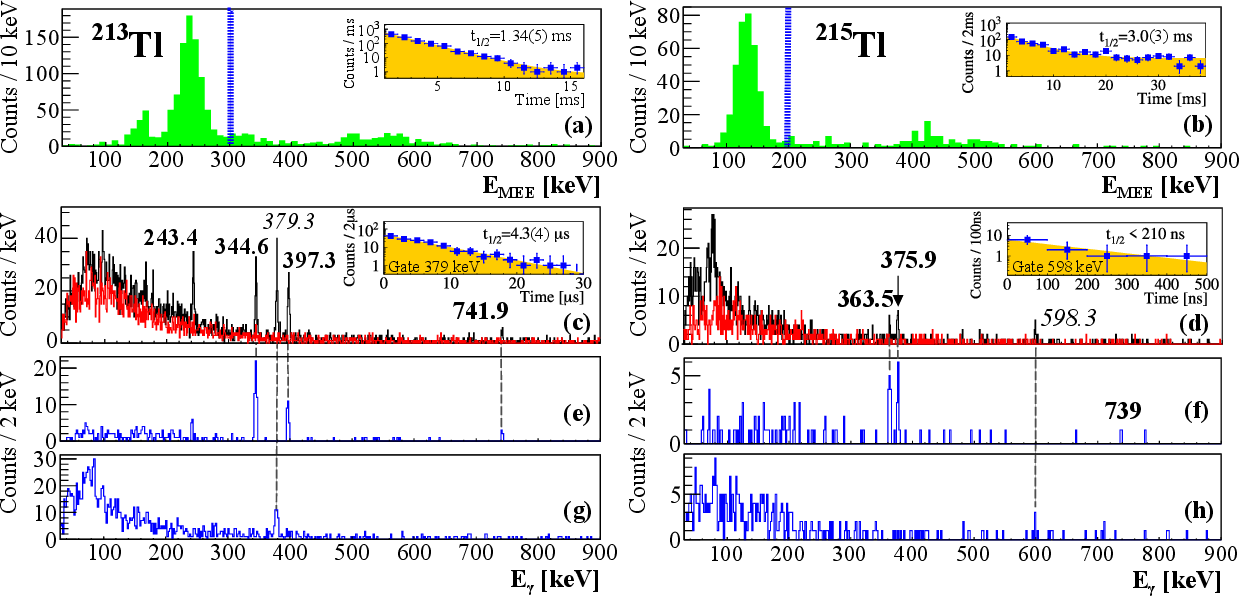}   
	\caption{(Color online) (a) and (b) MEE energy spectra of $^{213}$Tl and $^{215}$Tl, respectively. (c) and (d) Singles isomeric $\gamma$ spectra of $^{213}$Tl and $^{215}$Tl, respectively (black). For comparison, the background spectra of correlations between ions and preceding $\gamma$ rays are also shown (red). (e) and (f) $\gamma$-ray spectra in prompt-time coincidence with the MEEs at the left of the blue dotted lines in panels (a) and (b), respectively. (g) and (h) Spectra of $\gamma$ rays preceding the decay of the MEEs at the left of the blue dotted lines in panels (a) and (b), respectively. Inset panels in (a)\textendash(d) show time distributions of MEEs and preceding $\gamma$ rays [marked in italics in panels (c) and (d)]. Unbinned maximum likelihood fits to an exponential function plus a constant background are indicated in yellow.}
	\label{fig1}
\end{figure*}

Spectroscopic studies of heavy $N>126$ nuclei have been carried out for the first time at the RIBF facility, RIKEN Nishina Center. The cyclotron acceleration complex delivered a stable $^{238}$U beam at an energy of 345 MeV/nucleon. The beam intensity was the highest reached so far in a fragmentation facility, 70 pnA. The most exotic $N>126$ nuclei in the isotopic chains of $_{77}$Ir to $_{84}$Po were produced in fragmentation reactions of $^{238}$U colliding to a Be target 5-mm thick.  

Separation and identification of heavy nuclei are known to be a challenge in fragmentation facilities \cite{Mor11}. At RIKEN, unambiguous information on the mass-to-charge ratio $A/Q$ was obtained using thin degraders at the dispersive focal planes of the BigRIPS spectrometer  \cite{Fuk13}.  Identification in atomic number $Z$ was achieved using a novel telescope array consisting of four tilted Si detectors which provided accurate energy losses for the transmitted heavy fragments. The Si telescope was placed at the exit of the ZeroDegree spectrometer \cite{zerodegree}.

A total of about 4000 ions of $^{213}$Tl and 1000 ions of $^{215}$Tl were implanted in the active stopper WAS3ABi \cite{Nis12}. It consisted of a compact stack of four 1-mm thick Double-Sided Si Strip Detectors (DSSSD), each with $32\times32$ pixels. The implantation setup was completed by BRIKEN \cite{Tar17,Tol19}, a hybrid detection array  consisting of 140 $^{3}$He proportional counters and two segmented clover-type HPGe detectors. Due to severe damage of one of the two HPGe detectors in the experiment, only one of them was used for analysis, providing an absolute $\gamma$-ray efficiency of $\sim$1\% at 1408 keV. More details of the setup are provided in Ref. \cite{Tol19}.

Implanted nuclei, $\beta$ particles, and conversion electrons \textendash henceforth called Meitner-Ellis electrons or MEEs \cite{Han22}\textendash~were discriminated by the energy deposited in WAS3ABi. The active stopper also recorded the time, DSSSD and (X,Y) strip position for each of these events. The energy and time of $\gamma$ rays were registered by the  HPGe clover detector. These information were recorded on an event basis by independent data acquisition systems in order to reconstruct off-line the radioactive decay(s) of the implanted residues using correlations in position and time \cite{Tol19,Pho19,Wu22}.

Isomeric states were investigated in a broad dynamic time range exploiting two procedures. The first consists in building the time differences between implanted ions and subsequent $\gamma$ rays \cite{Ste11,Got12,Pho19,Wu22}. The shortest time differences are limited by the prompt flash produced by the implantation \cite{Ste11}, of approximately 200 ns. The second procedure exploits the production of MEEs in the internal decay of the isomer. In such cases, additional conditions on the discrete MEEs energies and (X,Y) strip distance between implanted ions and MEEs are applied. The shortest time difference for ion\textendash MEE correlations is fixed off-line to 500 $\mu$s to avoid the dead time caused by the electronic processing of implantation events.

Figures \ref{fig1}(a) and \ref{fig1}(b) show energy spectra of MEEs correlated with ions of $^{213}$Tl and $^{215}$Tl. The spatial correlations are restricted to the same DSSSD and pixel of implantation in WAS3ABi. Maximum ion\textendash MEE time differences of up to  40 ms are applied. Figs. \ref{fig1}(c) and \ref{fig1}(d) show $\gamma$ rays detected up to 6 ms after implantations of $^{213}$Tl and $^{215}$Tl, respectively. For the two nuclei, we find three $\gamma$ rays in 2-$\mu$s prompt-time delayed coincidence with the MEE peaks, see Figs. \ref{fig1}(e) and \ref{fig1}(f). The $\gamma$ transitions at 379.3(8) keV in $^{213}$Tl and 598.3(10) keV in $^{215}$Tl are found to decay up to 5 ms before the MEE [see Figs. \ref{fig1}(g) and \ref{fig1}(h)]. The time distributions of the MEEs and the preceding $\gamma$ rays, shown in the inset panels of Figs. \ref{fig1}(a)\textendash \ref{fig1}(d), reveal the existence of two isomeric states in each nucleus. Unbinned maximum likelihood fits including probability density functions of the isomer exponential decay and a constant background related to random correlations return half-lives of $t_{1/2} = 1.34(5)$ ms and $t_{1/2} = 4.3(4)$ $\mu$s for $^{213}$Tl, and  $t_{1/2} = 3.0(3)$ ms and  $t_{1/2} < 210$ ns for $^{215}$Tl. The present half-life of the 379.3(8)-keV $\gamma$ peak agrees within one standard deviation with the reported value of $t_{1/2}=4.1(5)$ $\mu$s \cite{Got19}, thus benchmarking the identification and correlation procedures used at RIBF.

\begin{figure}
	\centering
	\includegraphics[width=0.46\textwidth]{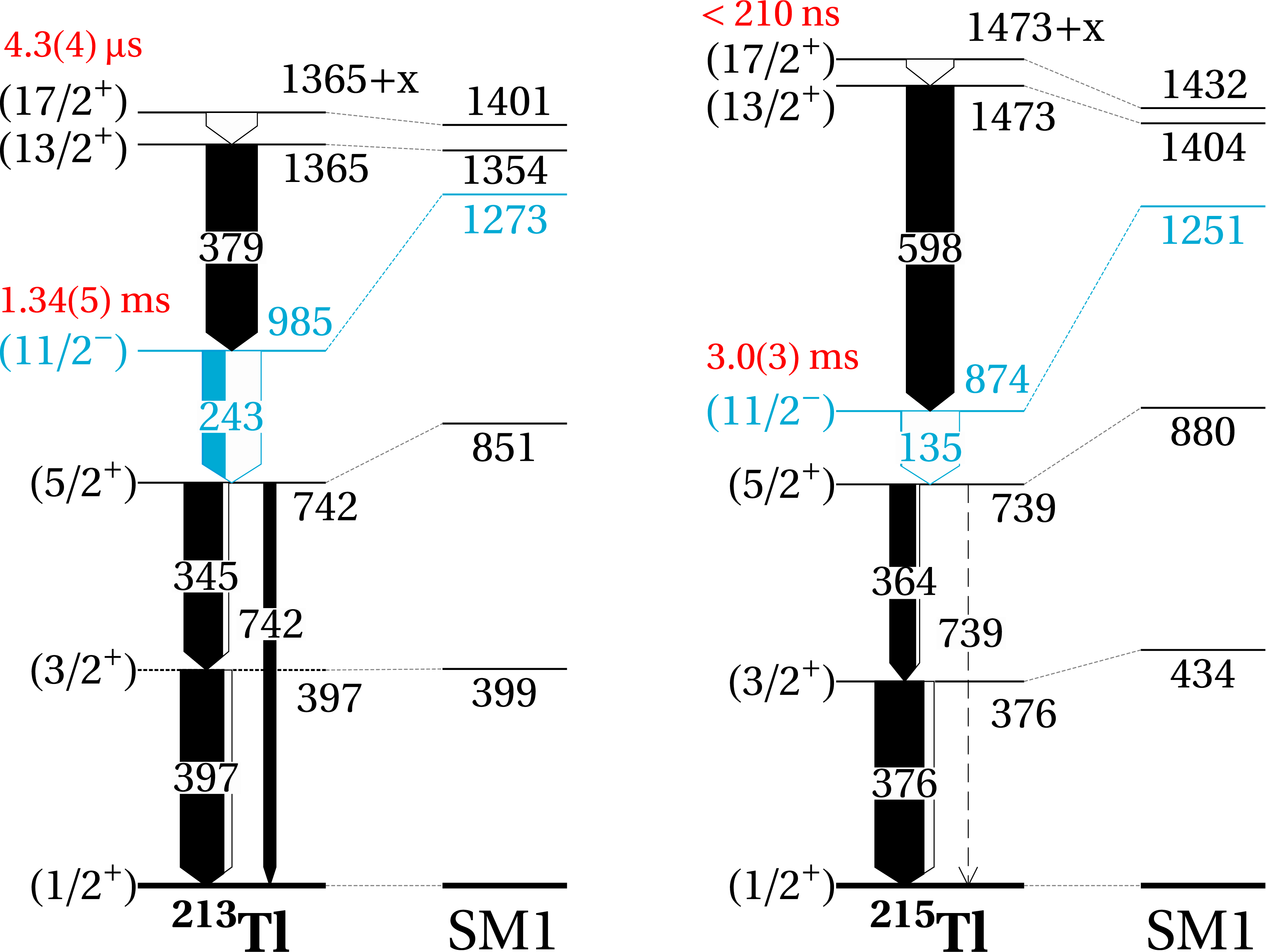}   
	\caption{(Color online) Isomeric decay schemes of $^{213}$Tl and $^{215}$Tl compared with shell-model calculations SM1 \cite{Yua22}. $11/2^-$ states and $E3$ transitions are indicated in blue, while isomeric half-lives are shown in red. Arrow widths are proportional to transition intensities.}
	\label{fig2}
\end{figure}

In Fig. \ref{fig2}, the isomeric level schemes of $^{213}$Tl and $^{215}$Tl are  compared to shell-model calculations (henceforth called SM1) using the code KSHELL \cite{Shi19}. The valence space consists of the proton orbitals $0g_{7/2}$, $1d_{5/2}$, $1d_{3/2}$, $2s_{1/2}$, $0h_{11/2}$ below $Z=82$ and the neutron orbitals $0i_{11/2}$, $1g_{9/2}$, $1g_{7/2}$, $2d_{5/2}$, $2d_{3/2}$, $3s_{1/2}$, $0j_{15/2}$ above $N=126$. The two-body matrix elements are based on the Kuo-Herling \cite{War91a,War91b} and VMU+LS \cite{Ots10,Ber77} interactions as described in Refs. \cite{Yua22,Liu23}. The single neutron and proton hole energies are taken from the experimental spectra of $^{209}$Pb and $^{207}$Tl, respectively. The transition probabilities are calculated using empirical effective charges of $e_{\nu}=0.8$ for neutrons and $e_{\pi}=1.8$ for protons \cite{Yua22}. For the most exotic $^{215}$Tl, we allow at most two neutrons exciting from the $\nu 1g_{9/2}$, $\nu 0i_{11/2}$, and $\nu 0j_{15/2}$ orbitals to the $\nu 1g_{7/2}$, $\nu 2d_{5/2}$, $\nu 2d_{3/2}$, and $\nu 3s_{1/2}$ ones.  No excitations across the $^{208}$Pb core are considered.

The experimental excitation spectra of the two nuclei are proposed based on the spectroscopic information of Fig. \ref{fig1}, $\gamma$-ray intensity balance, and comparison with SM1 calculations. They reveal alike structures, with a $(17/2^+)$ isomeric state decaying through the $(13/2^+)\rightarrow (11/2^-) \rightarrow (5/2^+) \rightarrow (3/2^+) \rightarrow (1/2^+)$ cascade. 
In the lighter $^{209}$Tl, the analogous $(17/2^+)$ long-lived level was previously attributed to the coupling of a $\pi s_{1/2}$ proton hole with the seniority  $\upsilon=2$, $(8^+)$ isomer in $^{210}$Pb \cite{Ald09,Amr17}. Similarly, the leading component of the $(17/2^+)$ states in $^{213}$Tl and $^{215}$Tl corresponds to the $\pi s_{1/2}^{-1} \otimes  \nu (g_{9/2})_{8^+}^n$ configuration. The isomeric nature of the newly observed states, hence, provides experimental evidence of good preservation of seniority in proton-hole-core coupled states well beyond $^{208}$Pb.  

Looking at Fig. \ref{fig2}, the $(17/2^+)$ levels decay via low-energy $E2$ transitions to the $(13/2^+)$ states, with predominant configuration $\pi s_{1/2}^{-1} \otimes \nu (g_{9/2})_{6^+}^n$. The absence of K$\alpha$ X rays in Figs. \ref{fig1}(c)\textendash\ref{fig1}(d) indicates that the energies of the $17/2^+ \rightarrow 13/2^+$ transitions in $^{213}$Tl and $^{215}$Tl are more likely below the binding energy of the Tl K electron, 85.5 keV. In $^{209}$Tl, the $(13/2^+)$ state decays to the $(9/2^+)$ level, with main $\pi s_{1/2}^{-1} \otimes \nu (g_{9/2})_{4^+}^{2}$ configuration, by a stretched $E2$ transition \cite{Ald09,Amr17}. For the more exotic $^{213}$Tl and $^{215}$Tl, the $(13/2^+)$ state no longer follows the seniority sequence but populates the $J^{\pi}=(11/2^-)$ level, mainly described by an unpaired proton hole in the $\pi h_{11/2}$ orbital, via an $E1$ decay. The $(11/2^-)$ state, which becomes isomeric, populates the yrast $(5/2^+)$ level arising from the $\pi s_{1/2}^{-1} \otimes \nu (g_{9/2})_{2^+}^{n}$ coupling. The calculated conversion coefficients  \cite{bricc} of the corresponding $E3$ transitions at 243.4(15) keV in $^{213}$Tl [$\alpha_{th}=1.48(2)$] and 135(5) keV in $^{215}$Tl [$\alpha_{th}=28(7)$] explain the pronounced Meitner-Ellis peaks in Figs. \ref{fig1}(a) and \ref{fig1}(b). The decay finally feeds the $J^{\pi}=(3/2^+)$ level, mainly described by a proton $\pi d_{3/2}^{-1}$ hole excitation, and the $J^{\pi}=(1/2^+)$ ground state, with leading configuration $\pi s_{1/2}^{-1}$.

\begin{figure*}
	\centering
	\includegraphics[width=\textwidth]{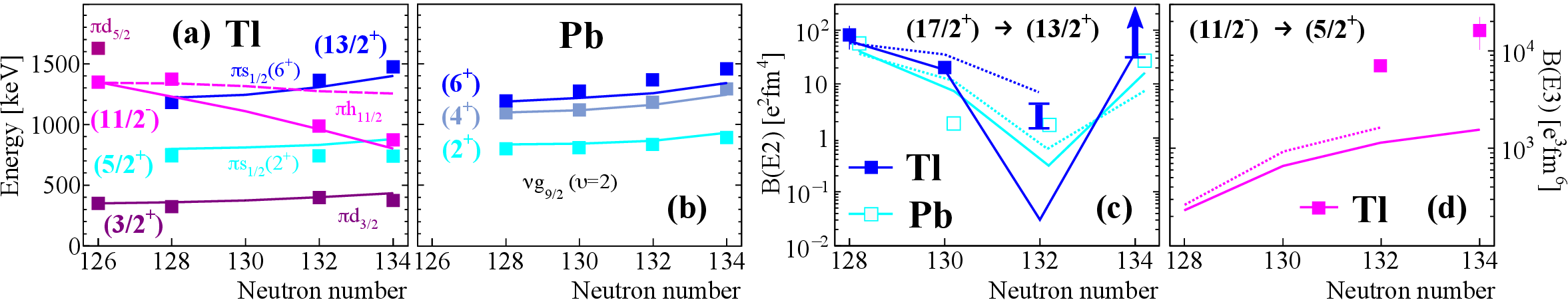}   
	\caption{(Color online) (a) Experimental levels of odd-mass Tl isotopes with $N\geq126$ (filled squares) compared to shell-model calculations (lines). States with leading single proton-hole configurations are shown in purple colors, while levels arising from proton-core couplings are marked in blue tones. (b) Same as (a) for the even-A Pb isotones. (c) Measured $B(E2;17/2^+\rightarrow 13/2^+)$ values for Tl (filled squares) compared to shell-model calculations. For $^{213}$Tl ($N=132$), the range of $B(E2)$ strengths for a transition of 20\textendash 80 keV is drawn. For $^{215}$Tl ($N=134$), the arrow  indicates the lower limit obtained from the isomeric lifetime measured in this work. Experimental $B(E2;8^+\rightarrow 6^+)$ strengths for the Pb isotones (empty squares) are also shown. (d) Same as (c) for $B(E3;11/2^-\rightarrow 5/2^+)$ strengths. In all panels, SM1 calculations are indicated in dashed lines, SM2 in solid lines, and SM3 in dotted lines. Experimental data for $^{207-211}$Tl and $^{210-216}$Pb are taken from Refs. \cite{Kon07,Ald09,Amr17,Got12,Got19}.}
	\label{fig3}
\end{figure*}

The excited states of odd-A Tl nuclei with two to eight neutrons above $N=126$, reported in this Letter and previous works \cite{Kon07,Amr17}, are shown by colored filled squares in Fig. \ref{fig3}(a). The unexpected fall of the $(11/2^-)$ state from 1369 keV in $^{209}$Tl to 985 keV in $^{213}$Tl and 874 keV in $^{215}$Tl provides the first stringent proof of shell evolution in the $Z<82,N>126$ quadrant beyond $^{208}$Pb. The SM1 calculations, shown in dashed magenta line, predict comparable excitation energies along the $\nu 1g_{9/2}$ orbital, failing to reproduce the lowering of the $\pi 0h_{11/2}$ proton-hole state in exotic Tl isotopes at increasing $N$. 

To argue the crucial role of the monopole interaction in reproducing the downward trend of the $J^{\pi}=(11/2^-)$ states, we first consider a constant shift of 300 keV in the $\pi 0h_{11/2}$ orbital SPE used in calculation SM1. The resulting $11/2^-$ energy in $^{213}$Tl agrees with experiment, but the calculation fails for $^{207,209}$Tl, predicting energies $\sim$300 keV lower. Yet one should note the ESPE is given by the addition of SPE and $n$$\cdot V_m$, where $n$ is the particle occupancy and $V_m$ is the monopole interaction. While the spherical doubly-magic SPE is constant, the ESPE changes as a function of $n$ owing to $V_m$. In the Tl isotopes, the reduction of $J^{\pi}=(11/2^-)$ excitation energies occurs as neutrons are added to the $\nu 1g_{9/2}$ orbital. Hence, a weakening of the proton-neutron monopole interaction between the $\pi 0h_{11/2}$ and $\nu 1g_{9/2}$ orbitals is more likely driving the observed lowering. To test the aforesaid hypothesis, we reduce the corresponding monopole matrix elements in SM1 by 60\%. This quantity is chosen to reproduce the energy of the $J^{\pi}=(11/2^-)$ level in $^{213}$Tl. We apply the truncation used in SM1 for $^{215}$Tl to calculate excitation energies and electromagnetic properties in all Tl nuclei. The theoretical results using the modified effective Hamiltonian and truncated model space are labeled as SM2 in Table \ref{table1} and shown in solid lines in Fig. \ref{fig3}. The new SM2 calculations do not only reproduce well the energy decrease of the $(11/2^-)$ levels, but the whole low energy spectra of $^{209-215}$Tl and $^{210-216}$Pb, as evinced in Figs. \ref{fig3}(a) and \ref{fig3}(b). Further, the lowering of the $(11/2^-)$ states by the monopole effect can be interpreted in the Nilsson mean-field model by a small increase in deformation due to an approach of the $\pi 2s_{1/2} ~[400]$ and $\pi 0h_{11/2}~ [505]$ orbitals of about the same energy.

Due to the particle-core mechanism \cite{Sha68}, the energies of the $(5/2^+)$ and $(13/2^+)$ states in odd-A Tl nuclei are similar to those of the seniority $(2^+)$ and $(6^+)$ levels in the $^{A+1}$Pb isotones, see Figs. \ref{fig3}(a) and  \ref{fig3}(b). The same situation is expected for the $B(E2, 17/2^+\rightarrow13/2^+)$ values, which should presumably follow the parabolic seniority behavior reported for the $(8^+)\rightarrow(6^+)$ transitions in Pb \cite{Got12}. Fig. \ref{fig3}(c) shows by filled squares the experimental $B(E2)$ strengths for Tl nuclei, where the reported $t_{1/2}=0.58(8)$-$\mu$s $\gamma$ ray at 144 keV in $^{211}$Tl \cite{Got19} is considered to de-excite the $(17/2^+)$ isomer. The analogous $B(E2)$ values in the $^{A+1}$Pb cores are also indicated by empty squares. The Tl nuclei show a breaking of symmetry around the middle of the $\nu 1g_{9/2}$ orbital, a behavior predicted for both Tl and Pb isotopes by SM2 (solid lines). This is in line with previous calculations using the Kuo-Herling and other realistic two-body interactions \cite{Got12}. The dissimilarity between the experimental transition strengths of $^{211}$Tl and $^{212}$Pb ($N=130$) is also significant, of about one order of magnitude. 
With our new data, we can further test the underlying residual interaction to understand the observed distortion from seniority in the Tl isotopes. To this aim, we have performed more realistic shell-model calculations in a larger model space (henceforth called SM3), including the five proton orbits below $Z=82$ and the $\pi 0h_{9/2}$, $\pi 1f_{7/2}$, and $\pi 0i_{13/2}$ orbitals above, and the $\nu 1g_{9/2}$, $\nu 0i_{11/2}$, and $\nu 0j_{15/2}$ orbitals above $N=126$. The calculations include one proton excitation across the $Z=82$ shell gap, allowing for the $\Delta j=2$ proton-core excitation $\pi s_{1/2}^{-1}(h_{11/2}^{-1}f_{7/2})\otimes(\nu g_{9/2}^n)$ in the Tl isotopes. This results in a general enhancement of the quadrupole collectivity, evinced by the dotted lines in Fig. \ref{fig3}(c). With SM3, the agreement between measured and calculated $B(E2)$ strengths improves significantly for the Tl isotopes. However, the experimental trend followed by the Pb nuclei is not reproduced, supporting the importance of neutron excitations across the $N=126$ shell gap in $Z=82$ nuclei \cite{Got12}.

The $B(E3; 11/2^- \rightarrow 5/2^+)$ strengths derived in the present work for $^{213}$Tl and $^{215}$Tl are shown by filled squares in Fig. \ref{fig3}(d). Comparison with calculations SM2 (solid line) and SM3 (dashed line) reveals discrepancies of about one order of magnitude between experiment and theory. According to SM2, the $E3$ transitions are mediated by admixtures of $\pi d_{5/2}^{-1}$ with the dominant $\pi s_{1/2}^{-1}\otimes2^+$ configuration in the $J^{\pi}=5/2^+_1$ states, resulting in sizable single-particle $\pi h_{11/2}\rightarrow \pi d_{5/2}$ $E3$ strength. These admixtures, which amount to 5\% in $^{213}$Tl and 8\% in $^{215}$Tl, explain the increasing trend of the calculated $B(E3)$, but not the excess of the experimental values. The missing strength might come from admixtures of the $\pi h_{11/2}^{-1}$ excitation with the $\pi d_{5/2}^{-1}\otimes3^-$ octupole coupling in the $11/2^-_1$ states. The $3^-$ collective octupole phonon in $^{208}$Pb \cite{Ham74} has been previously found to enhance the $E3$ rates between single-particle-like levels in neighboring nuclei \cite{For01,Ber20,Mor23}. The $B(E3)$ values reported here, in between the shell-model estimates and $B(E3)=33.8(6)$ W.u. for the $3^-$  phonon \cite{Mar07}, support this assumption.

\begin{table}
	\centering 
	\renewcommand{\arraystretch}{1.6}
	\renewcommand{\tabcolsep}{0.15 cm}
	\footnotesize
	\caption{Experimental excitation energies (in keV) of the $11/2^-_1$ states in odd-mass Tl isotopes compared to the shell-model calculations SM1 \cite{Yua22} and SM2, which use reduced $(\nu 1g_{9/2},\pi 0h_{11/2})$ matrix elements. New experimental information is indicated in bold. Previous data are taken from Refs. \cite{Kon07,Ell76}.} 
	\label{table1}
	
	\begin{tabular}{ c c c c c c }
		
		\hline
		\hline
		
		Nucleus & $^{207}$Tl & $^{209}$Tl & $^{211}$Tl & $^{213}$Tl & $^{215}$Tl \\
		
		\hline
	
		E$_{11/2^-_1}^{exp}$  & 1348.18(16) & 1369(10) &	\textendash     & \bf{985.3(19)}  & \bf{874(5)}  \\
		
		E$_{11/2^-_1}^{SM1}$ & 1349 & 1331 & 1306 & 1273 & 1251  \\
			
		E$_{11/2^-_1}^{SM2}$  & 1349 & 1230  & 1103 & 961 & 802   \\
		
		\hline
		\hline

	\end{tabular}
	
\end{table}

The SM3 calculations, which allow for one proton-core excitation to the $\pi 0h_{9/2}$, $\pi 1f_{7/2}$, and $\pi 0i_{13/2}$ orbitals above $Z=82$, predict slightly enhanced $B(E3)$ strengths than SM2. Though far from supplying a full treatment of the particle-octupole coupling effect in a shell-model framework, SM3 provides indications of what can be expected in a thorough computation of excitations across the shell gaps. Together with the $B(E2)$ results discussed before, our study evinces the importance of breaking the $^{208}$Pb core to understand the electromagnetic properties of $Z<82, N>126$ nuclei. 

In summary, first spectroscopy of the rarest $N>126$ Tl nuclei has been performed at RIKEN in a pioneering experiment. Excited isomeric structures with $J^{\pi}=(17/2^+)$ and $(11/2^-)$ were populated in $^{213}$Tl and $^{215}$Tl, providing spectral information not accessible in any other nuclear facility. 
The development of spin isomerism in the single proton-hole $\pi 0h_{11/2}$ states has been observed for the first time and attributed to a significant reduction of the monopole $(\nu 1g_{9/2},\pi 0h_{11/2})$ matrix elements, strongly supporting the crucial role of the monopole interaction \cite{Ots05} in shifting the $\pi 0h_{11/2}$ ESPE below $^{208}$Pb. The observed phenomenon resembles the reduction of the proton $\pi 0f_{7/2}-\pi 0f_{5/2}$ spin-orbit splitting by effect of the strong monopole interaction near $^{78}$Ni \cite{Tsu14}, which is boosted by the filling of the first $\nu 0g_{9/2}$ neutron shell \cite{Mor17,Sah17,Tan19}.
In parallel, the conservation of seniority has been proven through the excitation energy and $B(E2)$ of the $(17/2^+)$ isomers, which are interpreted as the coupling of a $\pi s_{1/2}$ proton hole to the $(8^+)$ seniority isomers in the  $^{A+1}$Pb isotones.
Such a study becomes unique in that no other isotopic chain provides a testing ground to investigate the impact of the particle-core mechanism \cite{Sha68} on the $B(E2)$ seniority cancellation rule across a complete $j$ orbital. 
The new results represent a first step to understanding how the high-$j$ orbitals evolve away from doubly-magic $^{208}$Pb, of special importance to model the $\beta$ decay towards the actinides in the \textit{r} process and pave the way to search for astromers beyond $N=126$, which can potentially influence the final abundance distribution of the heaviest elements.

We acknowledge the RI Beam Factory, operated by the RIKEN Nishina Center and CNS, University of Tokyo, for providing successful means to perform the experiment. Support of JSPS KAKENHI (Grants No. 14F04808, No. 17H06090, No. 25247045, 20H05648, 22H04946 and No. 19340074), Spanish Ministerio de Economia y Competitividad grants (FPA2011-06419, FPA2011-28770-C03-03, FPA2014-52823-C2-1-P, FPA2014-52823-C2-2-P, SEV-2014-0398, FPA2017-83946-C2-1-P, FPA2017-83946-C2-2-P, PID2019-104714GB-C21, PID2019-104714GB-C22), and Generalitat Valenciana, Conselleria de Innovaci\'on, Universidades, Ciencia y Sociedad Digital (CISEJI/2022/25, PROMETEO/2019/007, CIPROM/2022/9) is acknowledged. This research was sponsored by the Office of Nuclear Physics, U.S. Department of Energy under Awards No. DE-AC02-98CH10886 (BNL) and No. DE-AC02-06CH11357 (ANL), by the U.S. Department of Energy (DOE), Office of Science, Office of Nuclear Physics, under Grant No. DE-SC0020451, the Office of Nuclear Physics, U.S. Department of Energy under Award No. DE-FG02-96ER40983 (UTK) and DE-AC05-00OR22725 (ORNL), No. DE-SC0016988 (TTU) and by the National Nuclear Security Administration under the Stewardship Science Academic Alliances program through DOE Award No. DE-NA0002132, the National Science Foundation under Grants No. PHY-1430152 (JINA Center for the Evolution of the Elements), No. PHY-1565546 (NSCL), and No. PHY-1714153 (Central Michigan University), by Guangdong Major Project of Basic and Applied Basic Research under Grant No. 2021B0301030006, by the UK Science and Technology Facilities Council, by NKFIH (NN128072 and K147010), by European Commission FP7/EURATOM Contract No. 605203, by the UK Science and Technology Facilities Council Grant No. ST/N00244X/1, ST//P004598/1, and ST/V001027/1, by the National Research Foundation (NRF) in South Korea (No. 2016K1A3A7A09005575, No. 2015H1A2A1030275) and by the Natural Sciences and Engineering Research Council of Canada (NSERC) via the Discovery Grants SAPIN-2014-00028 and RG-PAS 462257-2014. TRIUMF receives federal funding via a contribution agreement with the National Research Council of Canada. T.T.Y. acknowledges support from Grant-in-Aid for JSPS Fellows Grant Numbers 23KJ0727; Forefront Physics and Mathematics Program to Drive Transformation (FoPM), a World-leading Innovative Graduate Study (WINGS) Program, the University of Tokyo; JSR Fellowship, CURIE, JSR-University of Tokyo Collaboration Hub; Teijin Kumura Scholarship, Teijin Scholarship Foundation; and Monbukagakusho Honors Scholarship, Japan Student Services Organization. R.Y. acknowledges support from JSPS KAKENHI 22K14053. 
A. A. Acknowledges partial support of the JSPS Invitational Fellowships for Research in Japan (ID: L1955). J.P. acknowledges support by the Institute for Basic Science (IBS-R031-D1).

\bibliographystyle{apsrev4-2}
\bibliography{bibliografia}

\end{document}